\documentclass{elsart}
\usepackage{graphicx}
\usepackage{color}
\usepackage{amsmath}
\usepackage{amssymb}
\usepackage{mathrsfs}         
\usepackage[all]{xy}
\newtheorem{open}{Question}
\usepackage{url}

\newcommand{\NP}[0]{\ensuremath{\mathsf{NP}}}

\newcommand{\NL}[0]{\ensuremath{\mathsf{NLogspace}}}
\newcommand{\Logspace}[0]{\ensuremath{\mathsf{L}}}

\newcommand{\Ptime}[0]{\ensuremath{\mathsf{P}}}
\newcommand{\Pspace}[0]{\ensuremath{\mathsf{Pspace}}}

\newcommand{\Hom}[0]{\ensuremath{\textsc{Hom}}}
\newcommand{\Comp}[0]{\ensuremath{\textsc{Comp}}}
\newcommand{\Ret}[0]{\ensuremath{\textsc{Ret}}}
\newcommand{\sHom}[0]{\ensuremath{\textsc{Sur\mbox{-}Hom}}}
\newcommand{\lHom}[0]{\ensuremath{\textsc{List\mbox{-}Hom}}}
\newcommand{\ntHom}[0]{\ensuremath{\textsc{Non\mbox{-}Triv\mbox{-}Hom}}}

\renewcommand{\phi}{\varphi}

\newcommand{\ignore}[1]{}

\begin{document}
\begin{frontmatter}

\title{The Complexity of Surjective Homomorphism Problems -- a Survey}
\author[lix]{Manuel Bodirsky\thanksref{fn1}}
\ead{bodirsky@lix.polytechnique.fr}
\author[prague]{Jan K\'ara\thanksref{fn2}}
\ead{jack@ucw.cz}
\author[ecs]{Barnaby Martin\corauthref{cor1}\thanksref{fn3}}
\ead{barnabymartin@gmail.com}
\corauth[cor1]{Corresponding author}
\thanks[fn1]{Supported by European Research Council under the European Community's Seventh Framework Programme (FP7/2007-2013 Grant Agreement no. 257039)}
\thanks[fn2]{The Institute for Theoretical Computer Science is supported as project IM0545 by the Ministry of Education of the Czech Republic.}
\thanks[fn3]{Supported by EPSRC grant EP/G020604/1.}

\address[lix]{CNRS/LIX, \'Ecole Polytechnique, France}
\address[prague]{Institute for Theoretical Computer Science, Charles University, Prague, Czech Republic}
\address[ecs]{Engineering and Computing Sciences, Durham University, U.K.}

\begin{abstract}
We survey known results about the complexity of surjective homomorphism problems, studied in the context of related problems in the literature such as list homomorphism, retraction and compaction. In comparison with these problems, surjective homomorphism problems seem to be harder to classify and we examine especially three concrete problems that have arisen from the literature, two of which remain of open complexity.
\end{abstract}

\end{frontmatter}

\section{Introduction}

The homomorphism problem, in its guise as constraint satisfaction, has numerous applications in various fields of computer science such as artificial intelligence and database theory. Many well-known problems in \NP\ may be formulated as homomorphism problems \cite{Jeavons} and in graph theory, where the problem is known as $\mathcal{H}$-colouring, there are results enough to fill a book \cite{HNBook}. The homomorphism problems we study ask whether a structure $\mathcal{A}$ has a homomorphism to a fixed structure $\mathcal{B}$ -- the template -- and a natural variant of this requires that the homomorphism be surjective. A homomorphism problem is trivial if the relations of $\mathcal{B}$ all contain the tuple $(b,\ldots,b)$, for some $b \in B$, and the matching surjective homomorphism problem is one of the most closely related problems for which this need no longer cause triviality.

Despite their naturalness, surjective homomorphism problems have attracted less attention in the literature than other homomorphism-related problems such as retraction and list homomorphism. In this paper we survey known results about surjective homomorphism problems, particularly in the context of their brethren homomorphism, compaction, retraction and list homomorphism problems. Perhaps the principal message of this survey is that surjective homomorphism problems seem to be very difficult to classify in terms of complexity -- that some of their number are possibly threshold cases, close to the boundary of intractability. We discuss why standard methods to prove easiness or hardness fail, and it is in this spirit that we present three concrete surjective problems, two of open complexity, that have arisen naturally in the literature. We would like to emphasise this distinction from the world of homomorphism problems, where for single explicitly given templates it is usually not difficult to classify the complexity of the corresponding problem.

Being a survey, it should not be necessary to address related work in the introduction as this should appear in the body of the paper. However, we mention here some work that is somehow similar but outside of the scope of this survey. Locally surjective homomorphism problems, also known as $\mathcal{H}$-role assignment, have been studied in \cite{FialaPaulusma,PimFialaPaulusma}. One of our central problems, $\sHom(\mathcal{C}^\mathrm{ref}_{4})$, is very closely related to a graph partition problem. There is a rich literature on graph partition and especially list partition problems. For list partition problems we mention particularly \cite{FederHellKleinMotwani} and \cite{Cameron-etal} (in the latter, another intriguing problem of open complexity appeared -- after six years it has just been classified as in \Ptime\ \cite{Stubborn}). The dichotomy for Boolean surjective homomorphism problems -- which will appear later -- has been used in the classification of a class of infinite-domain homomorphism problems related to phylogeny  problems \cite{BodirskyMueller}. Finally, asking that a homomorphism be surjective is a particular kind of global cardinality constraint. These have attracted much attention in the constraints community, and we mention in particular the complexity results and classifications of \cite{BulatovMarx} and \cite{CATS2008}.

The paper is organised as follows. In Section~\ref{sec:prelims} we introduce the problems that play a role in our survey together with the relationships that hold between them. In Section~\ref{sec:origins}, we consider the chronology in which various of these problems were considered as well as giving some basic results. We then examine why a full classification for surjective homomorphism problems is likely to be difficult (in fact all we do is infer this result from the equivalent results for retraction and compaction).
In Section~\ref{sec:cref4}, we introduce our first two problems $\sHom(\mathcal{C}_{6})$ and $\sHom(\mathcal{C}^\mathrm{ref}_{4})$ and look at recent work on $\sHom(\mathcal{C}^\mathrm{ref}_{4})$, culminating in its classification as \NP-complete. In Section~\ref{sec:no-rainbow}, we introduce our third problem of open complexity -- the $3$-no-rainbow-colouring problem -- and give some related results towards its classification. In particular, we introduce the idea of safe gadgets in reductions that do not artificially interfere with the condition of surjectivity. We then conclude the paper with some final remarks.

\section{Preliminaries}
\label{sec:prelims}

For some finite and relational signature $\sigma$, we consider only finite $\sigma$-structures $\mathcal{A}$, $\mathcal{B}$ \mbox{etc.} whose underlying domains we denote $A$, $B$ \mbox{etc.} of cardinality $|A|$, $|B|$ etc. A \emph{homomorphism} from $\mathcal{A}$ to $\mathcal{B}$ is a function $h:A \rightarrow B$ such that, for all $R \in \sigma$ of arity $i$, if $R(a_1,\ldots,a_i) \in \mathcal{A}$ then $R(h(a_1),\ldots,h(a_i)) \in \mathcal{B}$. 
The \emph{homomorphism problem} $\Hom(\mathcal{B})$ takes as input some finite $\mathcal{A}$ and asks whether there is a homomorphism from $\mathcal{A}$ to the fixed \emph{template} $\mathcal{B}$ (denoted $\mathcal{A} \rightarrow \mathcal{B}$). The \emph{surjective homomorphism problem} $\sHom(\mathcal{B})$ is defined similarly, only we insist that the homomorphism $h$ be surjective. It is easy to see that the problems $\Hom(\mathcal{B})$ and $\sHom(\mathcal{B})$ (and all the problems we will work with in this paper) are in \NP.
The problems $\Hom(\mathcal{B})$ span a broad subclass of \NP\ that may appear to form a microcosm. However, it is conjectured that these problems are always either in \Ptime\ or are \NP-complete \cite{FederVardi,JBK} -- a property that \NP\ itself does not have, assuming $\Ptime \neq \NP$ \cite{Ladner}. 

A \emph{digraph} is a structure with a single binary relation $E$. If $E$ is symmetric and antireflexive then $\mathcal{H}$ is a \emph{graph}. If $E$ is just symmetric, we call $\mathcal{H}$ a \emph{partially reflexive graph}.
We introduce several graphs that will play a role in our exposition. Let $[n]$ be the set $\{1,\ldots,n\}$. The cliques $\mathcal{K}_n$ and $\mathcal{K}^{\mathrm{ref}}_n$ each have domain $[n]$, with binary edge relations $E^{\mathcal{K}_n}:=\{ (i,j) : i\neq j\}$ and $E^{\mathcal{K}^{\mathrm{ref}}_n}:= [n]^2$, respectively. $\Hom(\mathcal{K}_n)$ is the \emph{graph $n$-colouring problem} which will appear in this survey many times. The cycles $\mathcal{C}_n$ and $\mathcal{C}^{\mathrm{ref}}_n$ each have domain $[n]$, with binary edge relations $E^{\mathcal{C}_n}:=\{ (i,j) : i-j \bmod n= \mbox{$1$ or $-1$} \}$ and $E^{\mathcal{C}^{\mathrm{ref}}_n}:=\{ (i,j) : i-j \bmod n= \mbox{$1$, $0$ or $-1$} \}$, respectively. The paths $\mathcal{P}_n$ and $\mathcal{P}^{\mathrm{ref}}_n$ each have domain $[n+1]$, with binary edge relations $E^{\mathcal{P}_n}:=\{ (i,j) : i-j = \mbox{$1$ or $-1$} \}$ and $E^{\mathcal{P}^{\mathrm{ref}}_n}:=\{ (i,j) : i-j = \mbox{$1$, $0$ or $-1$} \}$, respectively. The distance $d(i,j)$ between vertices $i,j \in G$ is the minimum length of a path in $\mathcal{G}$ between them. The \emph{diameter} of a graph is the maximum of $d(i,j)$ over all its vertices $i$ and $j$ (a disconnected graph has infinite diameter).
If $\mathcal{G}$ is an (antireflexive) graph, then its complement $\overline{\mathcal{G}}$ is defined over the same domain $G$, with edge set $E^{\overline{\mathcal{G}}}:=\{ (x,y) : (x,y) \notin E^{\mathcal{G}}, x \neq y\}$. In the context of graphs, the problem $\Hom(\mathcal{H})$ is usually known as the \emph{$\mathcal{H}$-colouring} problem.

There are several further problems, related to the homomorphism problem, that appear as though they might be relevant in studying the complexities of surjective homomorphism problems. Closely related is the \emph{non-trivial homomorphism problem} $\ntHom(\mathcal{B})$ which asks if there is a homomorphism from $\mathcal{A}$ to $\mathcal{B}$ that is not a constant function. Alongside the surjective homomorphism problem, this manifests as one of the more natural variants of homomorphism that may remain hard where that problem becomes easy. The \emph{list homomorphism problem} $\lHom(\mathcal{B})$ takes as input some $\mathcal{A}$ together with, for each $a \in A$, lists $L_a \subseteq B$, and asks whether there is a homomorphism $h$ from $\mathcal{A}$ to $\mathcal{B}$ such that $h(a) \in L_a$. List homomorphism is clearly a special case of homomorphism where the template is expanded by all possible unary relations (corresponding to all possible lists). Much is known about the complexity of list homomorphism problems. In \cite{FederHellHuang99} a dichotomy is proved for $\lHom(\mathcal{H})$ when $\mathcal{H}$ is a graph. Specifically, if the complement of $\mathcal{H}$ is a circular arc graph then $\lHom(\mathcal{H})$ is in \Ptime, otherwise $\lHom(\mathcal{H})$ is \NP-complete. In \cite{FederHell98}, a dichotomy is proved for $\lHom(\mathcal{H})$ when $\mathcal{H}$ is a reflexive graph. Specifically, if $\mathcal{H}$ is an interval graph then $\lHom(\mathcal{H})$ is in \Ptime, otherwise $\lHom(\mathcal{H})$ is \NP-complete (an \emph{interval graph} can be realised in the following fashion: the vertices are closed connected sub-intervals of $[0,1]$ and an edge connects two vertices iff the intervals overlap). A complete dichotomy was given for partially reflexive graphs in \cite{Biarc} and, finally, Bulatov gave a full dichotomy for list homomorphism in \cite{Conservative}.

The \emph{retraction problem} $\Ret(\mathcal{B})$ takes as input some $\mathcal{A}$, with $\mathcal{B}$ an induced substructure of $\mathcal{A}$, and asks whether there is a homomorphism $h:\mathcal{A} \rightarrow \mathcal{B}$ such that $h$ is the identity on $\mathcal{B}$. It is important that the copy of $\mathcal{B}$ is specified in $\mathcal{A}$; it can be that $\mathcal{B}$ appears twice as an induced substructure and there is a retraction from one of these instances but not to the other. The problem $\Ret(\mathcal{B})$ is easily seen to be logspace equivalent with the problem $\Hom(\mathcal{B}^\mathrm{c})$, where $\mathcal{B}^\mathrm{c}$ is $\mathcal{B}$ expanded with all constants (one identifies all elements assigned to the same constant and enforces the structure $\mathcal{B}$ on those constants). Thus, like list homomorphism, retraction problems are special cases of homomorphism problems (although they are at least as hard to fully classify -- see Theorem~\ref{thm:CSP=Ret}). In the context of graph problems, $\Hom(\mathcal{H}^\mathrm{c})$ is sometimes known as a \emph{precolouring problem}, due to the pre-assignment of the constants to the input.  

The \emph{compaction problem} is traditionally only defined on graphs, and we will first define it thus, as it may be generalised in more than one way. $\Comp(\mathcal{H})$ takes as input some $\mathcal{G}$ and asks whether there is a surjective homomorphism $h:\mathcal{G} \rightarrow \mathcal{H}$ such that $h$ is edge-surjective except possibly on self-loops. Formally, for all $h_1,h_2 \in H$ s.t. $h_1 \neq h_2$ and $E(h_1,h_2) \in \mathcal{H}$, there exists $g_1,g_2 \in G$ s.t. $E(g_1,g_2) \in \mathcal{G}$ and $h(g_1)=h_1$ and $h(g_2)=h_2$ ($E(g_1,g_2)$ is called a \emph{preimage} of $E(h_1,h_2)$). The stipulation therein that $h_1 \neq h_2$ appears rather unnatural (blame the graph-theorists!) and, other than this, one may say that this coincides with $\mathcal{H}$ being a \emph{homomorphic image} of $\mathcal{G}$ in the sense of \cite{HNBook}. The definition of compaction may be generalised to arbitrary signatures in at least two ways. Let $R$ be a relation of $\sigma$ of arity $i$. Firstly, we could insist that for every $b_1,\ldots,b_i \in B$ s.t. $b_1,\ldots,b_i$ \emph{are not all the same} and $R(b_1,\ldots,b_i) \in \mathcal{B}$, there exists $a_1,\ldots,a_i \in A$ s.t. $R(a_1,\ldots,a_i) \in \mathcal{A}$ and $h(a_1)=b_1$, \ldots, $h(a_i)=b_i$. Secondly, we could do likewise but with ``not all the same'' substituted by ``pairwise distinct''. We prefer the first generalisation and will henceforth stick with it.

\subsection{Relationship between the problems}

Let $\leq_\Logspace$ indicate many-to-one logspace reduction and $\leq^{\mathrm{Tur}}_\Ptime$ polynomial time Turing reduction.
\begin{prop}
\label{prop:reductions}
For finite $\mathcal{B}$ our problems sit in the following relationship.
\[ \Hom(\mathcal{B}) \leq_\Logspace \ntHom(\mathcal{B}) \leq^\mathrm{Tur}_\Ptime \sHom(\mathcal{B}) \ \ \ \ \ \ \ \ \ \ \ \ \ \  \]
\[  \ \ \ \ \ \ \ \ \ \ \ \ \ \ \leq^\mathrm{Tur}_\Ptime \Comp(\mathcal{B}) \leq^\mathrm{Tur}_\Ptime \Ret(\mathcal{B}) \leq_\Logspace \lHom(\mathcal{B}).\]
\end{prop}
\begin{pf}
It is a simple observation that both $\Hom(\mathcal{B}) \leq_\Logspace \ntHom(\mathcal{B})$ and $\Hom(\mathcal{B}) \leq_\Logspace \sHom(\mathcal{B})$. One may use the reduction maps $\mathcal{A} \mapsto \mathcal{A} \uplus \mathcal{K}_1$ (provided $|B|>1$) and $\mathcal{A} \mapsto \mathcal{A} \uplus \mathcal{B}$, respectively (the $\uplus$ indicates disjoint union; the designation $\mathcal{K}_1$ -- somewhat abused as the structures need not be digraphs -- means simply an element with empty relations). 

The reduction from $\ntHom(\mathcal{B})$ to $\sHom(\mathcal{B})$ proceeds as follows. From an input $\mathcal{A}$ for $\ntHom(\mathcal{B})$, we will consider each of the ways that two elements $a_1,a_2 \in A$ may be mapped to distinct $b_1,b_2$ in $B$ (somesuch must if there is to be a non-constant homomorphism). There are at most fewer than $|A|^2|B|^2$ many combinations $\chi$ to consider. For each one of these we build $\mathcal{A}_\chi$ by adjoining to $\mathcal{A}$ a copy of $\mathcal{B}$ with $a_1$ and $b_1$, as well as $a_2$ and $b_2$ identified. If one of these $\mathcal{A}_\chi$ surjectively maps to $\mathcal{B}$ then $A \subseteq A_\chi$ can not map to a constant element, for cardinality reasons (it would collapse $b_1$ and $b_2$ to the same element and one would be forced to surjectively map $|B|-1$ elements to $|B|$). Conversely, if $\mathcal{A}$ has a non-constant homomorphism to $\mathcal{B}$ then, by construction, one $\mathcal{A}_\chi$ will surjectively map to $\mathcal{B}$.

The reduction from $\sHom(\mathcal{B})$ to $\Comp(\mathcal{B})$ goes as follows. Let $\mathcal{B}$ be enumerated $\{b_1,\ldots,b_n\}$. $\mathcal{A} \in \sHom(\mathcal{B})$ iff for some $n$ constants $\{c_1,\ldots,c_n\} \subseteq A$ there is a homomorphism $h:\mathcal{A} \rightarrow \mathcal{B}$ s.t. $h(c_1)=b_1$, \ldots, $h(c_n)=b_n$. If $|A|=m$ then we consider all $m^n$ possible assignments $\chi$ for $c_1,\ldots,c_n$ in $A$, and we enforce the structure $\mathcal{B}$ on top of them; i.e. we build $\mathcal{A}_\chi$ from $\mathcal{A}$ by adding relations $R(c_{\lambda_1},\ldots,c_{\lambda_i}) \in \mathcal{A}_\chi$ if $R(b_{\lambda_1},\ldots,b_{\lambda_i}) \in \mathcal{B}$. We claim that $\mathcal{A} \in \sHom(\mathcal{B})$ iff, for some $\chi$, $\mathcal{A}_\chi \in \Comp(\mathcal{B})$. (Forwards.) Choose $\chi$ to witness surjectivity of the surjective homomorphism from $\mathcal{A}$ to $\mathcal{B}$. (Backwards.) A compaction is a fortiori a surjective homomorphism.

The reduction from $\Comp(\mathcal{B})$ to $\Ret(\mathcal{B})$ is discussed for graphs in \cite{VikasCSP} (see this for a more formal description). Given $\mathcal{A}$ of size $m$ as an input for $\Comp(\mathcal{B})$, one looks for some candidate set of preimages of the relations in $\mathcal{B}$. Since $\mathcal{B}$ is of finite size $n$, with say $\alpha$ relations of arity $\leq \beta$, this set of preimages is of size bound by $\alpha n^\beta$. Each preimage may mention $\beta$ elements of $\mathcal{A}$, so it follows that there are $\leq m^{\beta \alpha n^\beta}$ candidate sets of elements underlying these preimages in $\mathcal{A}$ (the important point being that this is a polynomial in $m$). When one has a candidate set, with the necessary preimage relations, one identifies elements as necessary and enforces the structure $\mathcal{B}$ on these elements, to produce a structure $\mathcal{A}'$. Finally, one of these $\mathcal{A}'$ will retract to $\mathcal{B}$ iff $\mathcal{A} \in \Comp(\mathcal{B})$. 

The reduction from $\Ret(\mathcal{B})$ to $\lHom(\mathcal{B})$ is trivial.
\end{pf}
We remark that the reduction from compaction to retraction breaks down on structures with infinite signatures. However, it is easy to see that there is a polynomial-time Turing reduction from surjective homomorphism to retraction that works even on structures with infinite signatures.

While $\Hom(\mathcal{B}) \leq_\Logspace \ntHom(\mathcal{B})$, the converse is unlikely (i.e. assuming $\Ptime \neq \NP$) to be true: take $\mathcal{B}:=\mathcal{K}_3 \uplus \mathcal{K}^{\mathrm{ref}}_1$. Clearly $\Hom(\mathcal{B})$ is trivial (one may map all vertices of the input to the $\mathcal{K}^{\mathrm{ref}}_1$. However, $\ntHom(\mathcal{B})$ is \NP-complete, as can be seen by reduction from \textsc{$3$-Colouring} \emph{for connected inputs only} (it is easy to see this remains \NP-complete; see, e.g., \cite{Papa}). One can use the reduction map $\mathcal{G} \mapsto \mathcal{G} \uplus \mathcal{K}^{\mathrm{ref}}_1$. Note that this same example also directly separates $\Hom(\mathcal{B})$ and $\sHom(\mathcal{B})$ (so long as the connected input for \textsc{$3$-Colouring} is of size $\geq 3$). Just as easily as $\Hom(\mathcal{B})$ and $\ntHom(\mathcal{B})$ were separated, so we may separate $\ntHom(\mathcal{B})$ and $\sHom(\mathcal{B})$ taking $\mathcal{H}:=\mathcal{K}_3 \uplus \mathcal{K}^{\mathrm{ref}}_1 \uplus \mathcal{K}^{\mathrm{ref}}_1$. Once again the reduction is from $\Hom(\mathcal{K}_3)$ and the map is $\mathcal{G} \mapsto \mathcal{G} \uplus \mathcal{K}^{\mathrm{ref}}_1 \uplus \mathcal{K}^{\mathrm{ref}}_1$.

We now establish how $\Ret(\mathcal{B})$ and $\lHom(\mathcal{B})$ may have differing complexities. Let $\mathcal{P}^{11100}_{4}$ be the $4$-path with self-loop on the first three vertices and not on the last two, i.e. with vertices $\{0,1,2,3,4\}$ and edge set $\{(x,y):|x-y|=1 \vee x,y=0 \vee x,y=1 \vee x,y=2 \}$). It is known that $\Ret(\mathcal{P}^{11100}_{4})$ is in \Ptime\ (see \cite{Pseudoforests}) while $\lHom(\mathcal{P}^{11100}_{4})$ is \NP-complete ($\mathcal{P}^{11100}_{4}$ has no conservative majority polymorphism; see \cite{Biarc}). No separation of the complexities of $\sHom(\mathcal{B})$, $\Comp(\mathcal{B})$ and $\Ret(\mathcal{B})$ is known. Furthermore, the following is a noted conjecture (this goes back to Winkler in 1988 for reflexive graphs; see Vikas's papers in the bibliography). 
\begin{conj}[Winkler, Vikas etc.]
For all graphs $\mathcal{H}$, $\Comp(\mathcal{H})$ and $\Ret(\mathcal{H})$ are polynomially Turing equivalent.
\end{conj}

\section{Origins of the problems}
\label{sec:origins}

Surjective homomorphisms problems are a natural generalisation of homomorphism problems, and it is in this context that we see our earliest classifications of surjective homomorphism problems. In the following we will not define what it means to be Horn, dual Horn, bijunctive or affine -- please see the text referenced.
\begin{thm}[\cite{CreignouHerbrard} (see \cite{Creignou})]
Let $\mathcal{B}$ be Boolean, \mbox{i.e.} on domain $\{0,1\}$. Then, if all relations of $\mathcal{B}$ are from one among Horn, dual Horn, bijunctive or affine, $\sHom(\mathcal{B})$ is in \Ptime. Otherwise, it is \NP-complete.
\end{thm}
This classification is very similar to the Boolean homomorphism dichotomy of Schaefer \cite{Schaefer}, except for the instances in which $\Hom(\mathcal{B})$ is trivial because the relations of $\mathcal{B}$ either all contain $(0,\ldots,0)$ or all contain $(1,\ldots,1)$. These latter cases, when the relations are not among Horn, dual Horn, bijunctive or affine, become hard in general for the surjective homomorphism problem. We note that in the Boolean case, the dichotomy for surjective homomorphism problems (between \Ptime\ and \NP-complete) coincides with the dichotomy for \emph{quantified constraint satisfaction problems} (between \Ptime\ and \Pspace-complete) \cite{Creignou}. 

Otherwise, and ancestrally, the most important problem from the perspective of surjective homomorphism, is list homomorphism. Since we are especially interested in graphs, we note specifically the work done by Feder, Hell and Huang \cite{FederHellHuang99,FederHell98}. Typically, their \NP-hardness results, proved in the context of list homomorphism, were immediately applied to give \NP-hardness for retraction, and then were modified by Vikas to give \NP-hardness for compaction. Two very important subcases were the \NP-hardness of $\lHom(\mathcal{C}_{2k})$ (for $k\geq 3$) \cite{FederHellHuang99} and the \NP-hardness of $\lHom(\mathcal{C}^{\mathrm{ref}}_{k})$ (for $k \geq 4$) \cite{FederHell98}. Although these appeared in different papers, the proofs are remarkably similar, and we will briefly review the proof in the case of $\mathcal{C}_6$.
\begin{prop}[\cite{FederHellHuang99}]
$\lHom(\mathcal{C}_{6})$ is \NP-complete ($\Ret(\mathcal{C}_{6})$ is \NP-complete).
\end{prop}
\begin{pf}[Sketch proof]
Membership of \NP\ is clear; \NP-hardness will be by reduction from \textsc{$3$-Colouring}. For a graph $\mathcal{G}$ as an input for \textsc{$3$-Colouring} we build a graph $\mathcal{G}''$ as an input for $\lHom(\mathcal{C}_{6})$ s.t. $\mathcal{G}$ is $3$-colourable iff $\mathcal{G}'' \in \lHom(\mathcal{C}_6)$. Firstly, we build $\mathcal{G}':=\mathcal{G} \uplus \mathcal{C}_6$. We set the lists $L_c$ of each $c \in \mathcal{C}_6$ to be $\{c\}$. All other lists will be set to the whole domain $C_6$. Finally, we build $\mathcal{G}''$ from $\mathcal{G}'$ by replacing every edge $E(x,y) \in \mathcal{G}$ with the gadget in Figure~\ref{fig:1} (which connects also to the fixed copy of $\mathcal{C}_6$ in $\mathcal{G}'$).
\begin{figure}
\begin{center}
\includegraphics{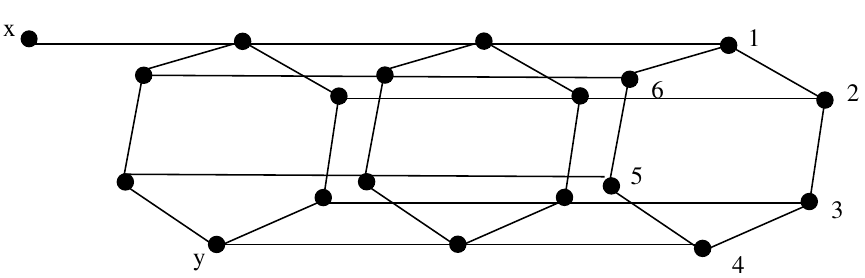}
\end{center}
\label{fig:1}
\caption{Gadget for \NP-hardness of $\lHom(\mathcal{C}_{6})$.}
\end{figure}
Note that, for each edge $E(x,y) \in \mathcal{G}$, fresh copies of all of the vertices drawn in the gadget are added, except of course for the fixed copy of $\mathcal{C}_6$. The vertices $2$, $4$ and $6$ represent the three colours -- it is not hard to see that if $x$ is mapped to one of these, then $y$ must be mapped to another.

Note that it is easy to see that the given proof is in fact a proof of \NP-hardness of $\Ret(\mathcal{C}_{6})$.
\end{pf}
\noindent The proofs for $\lHom(\mathcal{C}_{2k})$ and $\Ret(\mathcal{C}_{2k})$ (for $k\geq 4$) are very similar, involving reduction from $k$-colouring \cite{FederHellHuang99}. The proofs for $\lHom(\mathcal{C}^{\mathrm{ref}}_{k})$ and $\Ret(\mathcal{C}^{\mathrm{ref}}_{k})$ (for $k\geq 4$) are also very similar, involving again reduction from $k$-colouring \cite{FederHell98}.

Let $\mathcal{H}$ be either a bipartite graph or a reflexive graph. It is easy to see that $\Hom(\mathcal{H})$ is in \Logspace. In the latter case the problem is trivial; in the former the problem is either trivial or equivalent to \textsc{$2$-Colouring}, which is in \Logspace\ by the result of Reingold \cite{RheingoldJACM}. We have seen that there is a partially reflexive and disconnected graph that separates the complexities of homomorphism and surjective homomorphism. We will now see that we can go further.
\begin{prop}
There exists a bipartite graph $\mathcal{H}^\mathrm{bip}$ and a (connected) reflexive graph $\mathcal{H}^\mathrm{ref}$ such that $\sHom(\mathcal{H}^\mathrm{ref})$ and $\sHom(\mathcal{H}^\mathrm{bip})$ are both \NP-complete.
\end{prop}
\begin{pf}
We give the simple modification to the previous proof in the case of $\sHom(\mathcal{H}^\mathrm{bip})$. We assume, w.l.o.g., that the input $\mathcal{G}$ to \textsc{$3$-Colouring} has no isolated vertices and contains some edge. It follows that the diameter of $\mathcal{G}''$ in the previous proof is $8$ -- but in fact we will use a stronger condition than this which will become apparent. Set $\mathcal{H}^\mathrm{bip}$ to be $\mathcal{C}_6$ with distinct paths of length $3$ attached to each of its vertices (these paths will be known as tentacles). We use exactly the same reduction as before except that the fixed copy of $\mathcal{C}_6$ in $\mathcal{G}''$ becomes a fixed copy of $\mathcal{H}^\mathrm{bip}$ in $\mathcal{G}'''$ (other than this $\mathcal{G}'''$ is constructed as $\mathcal{G}''$). We claim that any surjective homomorphism $h$ from $\mathcal{G}'''$ to $\mathcal{H}^\mathrm{bip}$ must map $\mathcal{H}^\mathrm{bip} \subseteq \mathcal{G}'''$ to $\mathcal{H}^\mathrm{bip}$ by some automorphism, whereupon the result follows. In Figure~\ref{fig:2}, we depict the reduction gadget as appears in $\mathcal{G}'''$.
\begin{figure}
\begin{center}
\includegraphics{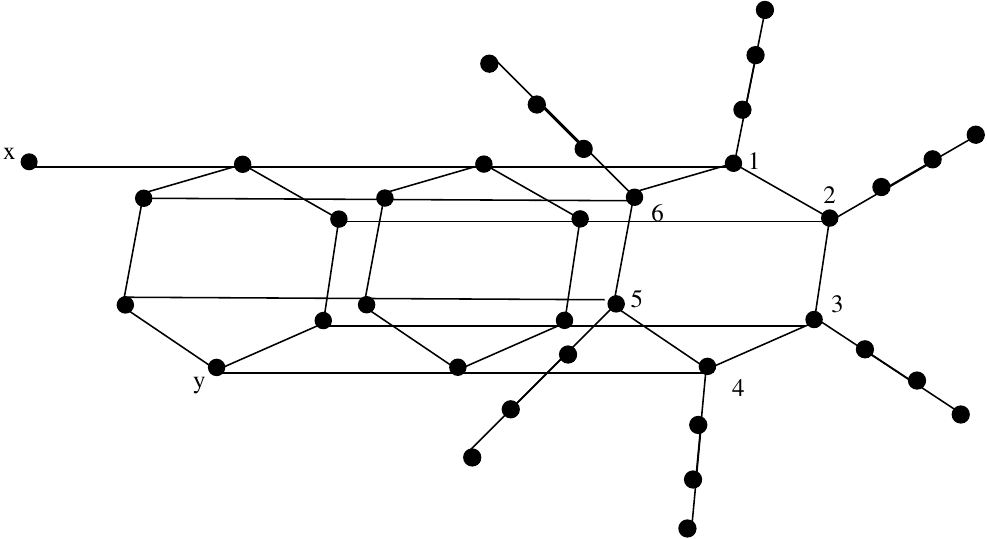}
\end{center}
\label{fig:2}
\caption{Gadget for \NP-hardness of $\sHom(\mathcal{H}^\mathrm{bip})$.}
\end{figure}
It remains to prove the claim. The diameter of $\mathcal{G}''$ was $8$; the diameter of $\mathcal{G}'''$ is in fact $9$. If $\mathcal{H}^\mathrm{bip} \subseteq \mathcal{G}'''$ is not mapped to $\mathcal{H}^\mathrm{bip}$ according to some automorphism of $\mathcal{H}^\mathrm{bip}$ then the homomorphic image $h(\mathcal{C}_6)$ in $\mathcal{H}^\mathrm{bip}$ for $\mathcal{C}_6 \subseteq \mathcal{H}^\mathrm{bip} \subseteq \mathcal{G}'''$ must be either $\mathcal{P}_3$, $\mathcal{P}_2$ or $\mathcal{P}_1$. We give the argument for $\mathcal{P}_3$; the same argument works in the other cases, which are in fact easier. Furthermore, we will assume that $h(\mathcal{C}_6)=\mathcal{P}_3 \subseteq \mathcal{C}_6 \subseteq \mathcal{H}^\mathrm{bip}$; the same argument works in the other cases -- \mbox{i.e.} if $h(\mathcal{C}_6)$ travels up a tentacle of $\mathcal{H}^\mathrm{bip}$ -- which are easier. Suppose, \mbox{w.l.o.g.} that $h(\mathcal{C}_6):=\{1,2,3,4\}$. It is easy to see that it is not possible for $h(\mathcal{H}^\mathrm{bip})$ to cover either of the farthest vertices on the tentacles connected to either of $\{5,6\} \subseteq \mathcal{C}_6 \subseteq \mathcal{H}^\mathrm{bip}$. It is now simple to see that no part of $\mathcal{G}'''$ can cover these in a homomorphism extending $h$ and we are done.

For $\mathcal{H}^\mathrm{ref}$ one may take $\mathcal{C}^{\mathrm{ref}}_{4}$ with distinct reflexive paths of length $2$ added to each vertex. The proof is similar to the above. 
\end{pf}   
\noindent We remark that one could have used $\mathcal{C}_6$ with paths of length $2$ added in the previous proof (though the argument is a little harder). One could also take $\mathcal{C}_6$ with only four paths of length $3$, so long as the vertices of $\mathcal{C}_6$ without paths are not adjacent.

Following these results on retraction, Vikas came along to prove matching results about compaction, as well as an equal classification for these two problems on partially reflexive graphs of size at most $4$.
\begin{thm}[\cite{VikasIrreflexive,VikasReflexive,VikasOther}]
$\Comp(\mathcal{C}_{2k})$ (for $k \geq 3$) and $\Comp(\mathcal{C}^\mathrm{ref}_{k})$ (for $k\geq 4$) are \NP-complete. If $\mathcal{H}$ is a partially reflexive graph s.t. $|H|\leq 4$, then $\Ret(\mathcal{H})$ and $\Comp(\mathcal{H})$ are polynomial time Turing equivalent.
\end{thm}

\subsection{Further simple classifications}

The chain of reductions in Proposition~\ref{prop:reductions} affords a simple answer to some classifications. A \emph{core} is a structure all of whose endomorphisms are automorphisms (an endomorphism of a structure is a homomorphism from it to itself).
\begin{lem}
If $\mathcal{B}$ is a core then $\Hom(\mathcal{B})$ and $\sHom(\mathcal{B})$ are polynomial-time Turing equivalent.
\end{lem}
\begin{pf}
It is well-known in this case that $\Hom(\mathcal{B})$ and $\Hom(\mathcal{B}^\mathrm{c}):=\Ret(\mathcal{B})$ are logspace equivalent \cite{JBK}. The result follows from Proposition~\ref{prop:reductions}.
\end{pf}
\begin{cor}
Let $\mathcal{H}$ be a semicomplete digraph. If $\mathcal{H}$ contains at most one cycle then $\sHom(\mathcal{H})$ is in \Ptime, otherwise $\sHom(\mathcal{H})$ is \NP-complete.
\end{cor}
\begin{pf}
Semicomplete digraphs are cores, therefore we have the same classification as for the homomorphism problems \cite{Semicomplete} (it is also not so hard to prove the polynomial cases directly).
\end{pf}

\subsection{Difficulty of a full classification}
\label{sec:diff}

For a long time a link has been known between the complexity classifications of homomorphism and retraction problems. In one of the deepest papers written on homomorphism problems, Feder and Vardi proved the following.
\begin{thm}[\cite{FederVardi}]
\label{thm:CSP=Ret}
For every structure $\mathcal{B}$ there exists a bipartite (antireflexive) graph $\mathcal{H}$ such that $\Hom(\mathcal{B})$ and $\Ret(\mathcal{H})$ are polynomial-time equivalent.
\end{thm}
\noindent Armed with this result Feder and Hell were able to produce a similar result in the reflexive case.
\begin{thm}[\cite{FederHell98}]
For every structure $\mathcal{B}$ there exists a reflexive graph $\mathcal{H}$ s.t. $\Hom(\mathcal{B})$ and $\Ret(\mathcal{H})$ are polynomial-time equivalent.
\end{thm}
\noindent We already know from the introduction that retraction problems are special instances of homomorphism problems. These theorems each provide a kind of inverse. A full classification for retraction problems is as difficult as a full classification for homomorphism problems (and for the former we may even restrict to either of the classes of bipartite or reflexive graphs). 

Variants of the above theorems are given by Vikas for compaction \cite{VikasCSP}. For any bipartite (respectively, reflexive) $\mathcal{H}$ he produces a $\mathcal{H}'$ such that $\Ret(\mathcal{H})$ and $\Comp(\mathcal{H}')$ are polynomial-time equivalent. The method he employs uses paths rather like the tentacles of the Section~\ref{sec:origins}, only in a slightly more sophisticated manner. The intuition behind his proof is simple, but the proof itself is laborious and technical. The interesting thing from our perspective is that his reduction is actually also a reduction between $\Ret(\mathcal{H})$ and $\sHom(\mathcal{H}')$.
\begin{thm}[\cite{VikasCSP}]
For every structure $\mathcal{B}$ there exists a bipartite graph (respectively, reflexive graph) $\mathcal{H}$ s.t. $\Hom(\mathcal{B})$ and $\Comp(\mathcal{H})$ are polynomial-time equivalent. For every structure $\mathcal{B}$ there exists a bipartite graph (respectively, reflexive graph) $\mathcal{H}$ s.t. $\Hom(\mathcal{B})$ and $\sHom(\mathcal{H})$ are polynomial-time equivalent.
\end{thm}

\section{A renaissance in foresting}
\label{sec:cref4}

Work is ongoing for retraction and surjective homomorphism problems. A \emph{pseudoforest} is the disjoint union of \emph{pseudotrees} -- graphs containing at most one cycle. The following is a recent result.
\begin{thm}[\cite{Pseudoforests}]
\label{pseudo}
If $\mathcal{H}$ is a partially reflexive pseudoforest, then either $\Ret(\mathcal{H})$ is in \Ptime, or it is \NP-complete. The \NP-complete cases occur when either the looped vertices in a connected component of $\mathcal{H}$ induce a disconnected graph, or $\mathcal{H}$ contains a partially reflexive cycle on $5$-vertices, or $\mathcal{H}$ contains a reflexive cycle on $4$-vertices, or $\mathcal{H}$ contains an antireflexive cycle on $3$-vertices. 
\end{thm}
\noindent A part of this dichotomy has been proved for the surjective homomorphism problem even more recently.
\begin{thm}[\cite{GolovachPaulusmaSong}]
\label{GPS}
Let $\mathcal{T}$ be a partially reflexive forest. Then, if the vertices in each tree $\mathcal{T}$ with a self-loop induce a subtree (connected component) of $\mathcal{T}$, $\sHom(\mathcal{T})$ is in \Logspace. Otherwise, it is \NP-complete.\footnote{In \cite{GolovachPaulusmaSong} the dichotomy is primarily given for partially reflexive trees (the improvement to forests is only mentioned at the end of the paper). Also, the tractable cases are given as being in \Ptime; but the proven first-order expressibility demonstrates that \Logspace\ membership can be derived.}
\end{thm}
\noindent Ideally one would like to extend Theorem~\ref{GPS} to something like Theorem~\ref{pseudo}, but this would entail settling the complexities of $\sHom(\mathcal{C}_{2k})$ (for $k \geq 3$) and $\sHom(\mathcal{C}^\mathrm{ref}_{k})$ (for $k\geq 4$). The base cases for these sequences we give specifically.
\begin{open}
\label{open:c6}
What is the complexity of $\sHom(\mathcal{C}_{6})$.
\end{open}
\begin{open}
\label{open:cref4}
What is the complexity of $\sHom(\mathcal{C}^\mathrm{ref}_{4})$.
\end{open}
\noindent We note that both $\ntHom(\mathcal{C}_{6})$ and $\ntHom(\mathcal{C}^\mathrm{ref}_{4})$ are in \Logspace\ (the former asks for a proper $2$-colouring and the latter asks that the input be of size $\geq 2$). This gives further evidence to the nature of surjective homomorphism problems being threshold, bordered closely on each side by problems of known and different complexity.

As Vikas extended the hardness for retraction to hardness for compaction, why can we not do likewise for surjective homomorphism? It is easy to see that $\Comp(\mathcal{C}_{6})$ and $\sHom(\mathcal{C}_{6})$ coincide for inputs of diameter $\leq 4$. In a similar manner $\Comp(\mathcal{C}^\mathrm{ref}_{4})$ and $\sHom(\mathcal{C}^\mathrm{ref}_{4})$ coincide for inputs of diameter $\leq 2$ (for inputs of diameter $\geq 3$, $\sHom(\mathcal{C}^\mathrm{ref}_{4})$ is trivially true while $\Comp(\mathcal{C}^\mathrm{ref}_{4})$ may not be). Yet Vikas's reduction from $\Ret(\mathcal{C}_{6})$ to $\Comp(\mathcal{C}_{6})$ (respectively, $\Ret(\mathcal{C}^\mathrm{ref}_{4})$ to $\Comp(\mathcal{C}^\mathrm{ref}_{4})$) generates a graph of diameter $5$ (respectively, $3$). It seems not to be possible to preserve his gadgets while bringing down this diameter. It follows that one may rephrase Question~\ref{open:c6} (resp., \ref{open:cref4}) as what is the complexity of $\Comp(\mathcal{C}_{6})$ (resp., $\sHom(\mathcal{C}^\mathrm{ref}_{4})$) on inputs of diameter $\leq 4$ (resp., $\leq 2$).

The alternative, of course, is that both $\sHom(\mathcal{C}_{6})$ and $\sHom(\mathcal{C}^\mathrm{ref}_{4})$ are in \Ptime. A standard method to prove that a problem $\lHom(\mathcal{H})$ is in \Ptime\ is reduction to \textsc{$2$-Sat} \cite{FederHellHuang99,FederHell98} (the list homomorphism partially reflexive graph cases that are in \Ptime\ are actually all in \NL). Repeated efforts to do this have so far failed. 

We note here that several interesting graph complexity classifications are unclear for inputs of restricted diameter. For example, it is easy to see that \textsc{$k$-Colouring} is \NP-complete for inputs of diameter $\leq 2$ when $k\geq 4$. Also, \textsc{$3$-Colouring} is \NP-complete for inputs of diameter $\leq 4$ (see the standard proof in \cite{Papa}). But the case of \textsc{$3$-Colouring} for inputs of diameter $\leq 3$ (also diameter $\leq 2$!) remains open \cite{3ColD23}.

\subsection{Recent work on $\sHom(\mathcal{C}^\mathrm{ref}_{4})$}

The problem $\sHom(\mathcal{C}^\mathrm{ref}_{4})$ has attracted interest from the graph theory community, where it is known by a variety of names, but especially \emph{disconnected cut} (\cite{ItoKaminskiPaulusmaThilikos,Separators}; one mention in \cite{FleischnerMujuniPaulusmaSzeider}) and \emph{$2\mathcal{K}_2$-partition} (\cite{DantasMaffraySilva,CookDantas,External4}). The former problem involves finding a vertex cutset in $\mathcal{G}$ (a set whose removal results in a disconnected graph) such that the cutset itself is disconnected. It is clear that this coincides with $\sHom(\mathcal{C}^\mathrm{ref}_{4})$ (diagonally opposite vertices form the cutset). It is now known that this problem is tractable on several graph classes (see the reference for further definitions).
\begin{thm}[\cite{FleischnerMujuniPaulusmaSzeider}]
$\sHom(\mathcal{C}^\mathrm{ref}_{4})$ is in \Ptime\ for

\vspace{0.3cm}
\begin{tabular}{ll}
$\bullet$ graphs of diameter not equal to $2$ & $\bullet$ graphs of bounded maximum degree \\
$\bullet$ graphs not locally connected & $\bullet$ triangle-free graphs \\
$\bullet$ graphs with a dominating edge \\
\end{tabular}
\end{thm}
\noindent These results were extended in a further paper.
\begin{thm}[\cite{ItoKaminskiPaulusmaThilikos}]
$\sHom(\mathcal{C}^\mathrm{ref}_{4})$ is in \Ptime\ for 

\vspace{0.3cm}
\begin{tabular}{ll}
$\bullet$ apex-minor-free graphs & $\bullet$ connected chordal graphs
\end{tabular}
\end{thm}
\noindent We note that the class of apex-minor-free graphs includes all planar graphs. One particularly interesting case left open in this last paper is whether $\sHom(\mathcal{C}^\mathrm{ref}_{4})$ is in \Ptime\ for claw-free graphs (graphs without an induced copy isomorphic to vertices $\{0,1,2,3\}$ and edges $\{(x,y):x=0 \vee y=0\}\setminus \{(0,0)\}$).

There is another problem \emph{$2\mathcal{K}_2$-partition}, to partition the vertices of a graph into four nonempty classes $A,B,C$ and $D$ such that every vertex in $A$ is adjacent to every vertex in $B$ and every vertex in $C$ is adjacent to every vertex in $D$. $2\mathcal{K}_2$-partition is logspace-equivalent to $\sHom(\mathcal{C}^{\mathrm{ref}}_4)$ under the complement reduction map $\mathcal{G} \mapsto \overline{\mathcal{G}}$. $2\mathcal{K}_2$-partition is actually more the motivating problem in \cite{FleischnerMujuniPaulusmaSzeider}, where it is known as \emph{$2$-biclique vertex-cover}. In \cite{DantasH}, $2\mathcal{K}_2$-partition is the only one of a family of problems whose complexity is left open (this corresponds to the complexity of $\sHom(\mathcal{H})$ being known for all graphs $\mathcal{H}$ of size $4$ other than $\mathcal{C}^\mathrm{ref}_{4}$). $2\mathcal{K}_2$-partition is considered in \cite{DantasMaffraySilva,CookDantas}, where the following results appear.
\begin{thm}[\cite{DantasMaffraySilva}]
$\sHom(\mathcal{C}^\mathrm{ref}_{4})$ is in \Ptime\ when the complement of the input is in one of the following classes.

\vspace{0.3cm}
\begin{tabular}{ll}
$\bullet$ $\mathcal{K}_4$-free graphs & $\bullet$ diamond-free graphs \\
$\bullet$ planar graphs & $\bullet$ graphs of bounded treewidth \\
$\bullet$ claw-free graphs & $\bullet$ $(\mathcal{C}_5,\mathcal{P}_5)$-free graphs \\
$\bullet$ graphs with few $\mathcal{P}_4$s
\end{tabular}
\end{thm}
\begin{thm}[\cite{CookDantas}]
$\sHom(\mathcal{C}^\mathrm{ref}_{4})$ is in \Ptime\ when the complement of the input is in one of the following classes.

\vspace{0.3cm}
\begin{tabular}{ll}
$\bullet$ $\mathcal{C}_4$-free graphs & $\bullet$ circular arc graphs \\
$\bullet$ spiders & $\bullet$ $\mathcal{P}_4$-sparse graphs \\
$\bullet$ bipartite graphs \\
\end{tabular}
\end{thm}
\noindent Indeed, $2\mathcal{K}_2$-partition is considered sufficiently important to raise its own complexity class \emph{2K2-hard} in \cite{External4} (echoed in \cite{Versus}). The complexity of this problem, \mbox{a.k.a.} $\sHom(\mathcal{C}^\mathrm{ref}_{4})$ -- Question 2 -- was finally settled as being \NP-complete in \cite{BarnyDaniel}. Question 1, the complexity of $\sHom(\mathcal{C}_{6})$, remains open.

\section{The no-rainbow-colouring problem}
\label{sec:no-rainbow}

\begin{figure}
\begin{center}
\begin{picture}(0,0)%
\includegraphics{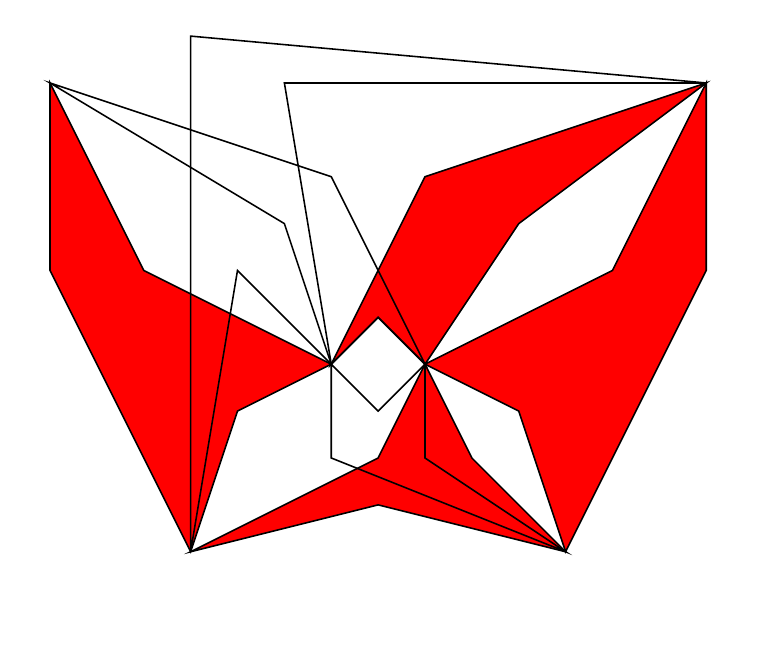}%
\end{picture}%
%
%
\setlength{\unitlength}{3947sp}%
\begingroup\makeatletter\ifx\SetFigFont\undefined%
\gdef\SetFigFont#1#2#3#4#5{%
  \reset@font\fontsize{#1}{#2pt}%
  \fontfamily{#3}\fontseries{#4}\fontshape{#5}%
  \selectfont}%
\fi\endgroup%
\begin{picture}(3630,3151)(661,-3279)
\put(1351,-3211){\makebox(0,0)[lb]{\smash{{\SetFigFont{12}{14.4}{\rmdefault}{\mddefault}{\updefault}{\color[rgb]{0,0,0}$0$}%
}}}}
\put(3601,-3211){\makebox(0,0)[lb]{\smash{{\SetFigFont{12}{14.4}{\rmdefault}{\mddefault}{\updefault}{\color[rgb]{0,0,0}$1$}%
}}}}
\put(4276,-286){\makebox(0,0)[lb]{\smash{{\SetFigFont{12}{14.4}{\rmdefault}{\mddefault}{\updefault}{\color[rgb]{0,0,0}$x$}%
}}}}
\put(676,-286){\makebox(0,0)[lb]{\smash{{\SetFigFont{12}{14.4}{\rmdefault}{\mddefault}{\updefault}{\color[rgb]{0,0,0}$y$}%
}}}}
\put(2139,-2086){\makebox(0,0)[lb]{\smash{{\SetFigFont{12}{14.4}{\rmdefault}{\mddefault}{\updefault}{\color[rgb]{0,0,0}$l$}%
}}}}
\put(2876,-1699){\makebox(0,0)[lb]{\smash{{\SetFigFont{12}{14.4}{\rmdefault}{\mddefault}{\updefault}{\color[rgb]{0,0,0}$r$}%
}}}}
\end{picture}%
\end{center}
\caption{A stylised depiction of the gadget for the reduction of $\Ret(\mathcal{N}_3)$ to $\sHom(\mathcal{N}_3;0,1)$. The gadget comprises six vertices and seven constraints -- four grey and three blank. The grey constraints are $\{y,0,l\}$, $\{l,r,x\}$, $\{r,1,x\}$, $\{0,r,1\}$ and the blank constraints are $\{y,l,r\}$, $\{0,l,x\}$, $\{l,r,1\}$.}
\label{fig:pcspn}
\end{figure}

A \emph{$k$-uniform hypergraph} may be seen as a structure with a single $k$-ary relation (which would usually be seen as being closed under permutations of position of its entries, and having always distinct entries in its tuples -- but this will not be important for us). The \emph{$k$-no-rainbow-colouring} problem takes as input a $k$-uniform hypergraph and asks whether there is a colouring of its vertices with all $k$ colours such that no hyperedge (tuple in the relation) attains all colours (is \emph{rainbow coloured}). This problem surfaced in the graph theory literature. (It is strict colouring of co-hypergraphs -- see \cite{ColoringMixedHypertrees}. In that paper the problem is conjectured to be in \Ptime\ -- though actually only a proof of \NP-hardness would settle one of the paper's outstanding cases.) It is simple to give a polynomial algorithm for the case $k=2$, but, for all higher $k$ the complexity is open. Henceforth, we will consider only the case  $k=3$. The $3$-no-rainbow-colouring problem can easily be cast as a surjective homomorphism problem with structure $\mathcal{N}$ on domain $N:=\{0,1,2\}$ and with ternary relation 
\[ R^\mathcal{N}:=\{0,1,2\}^3 \setminus \{(x,y,z)\ :\ x,y,z \mbox{ distinct} \}.\]
\begin{open}
\label{open:N}
What is the complexity of $\sHom(\mathcal{N})$.
\end{open}
It is not hard to see that $\Ret(\mathcal{N})=\Hom(\mathcal{N}^\mathrm{c})$ is \NP-complete. One may use Bulatov's classification \cite{Bulatov}, but we will give a direct proof. We will use the notation $\mathcal{N}^\mathrm{c}=(\mathcal{N};0,1,2)$ which shows the constants directly, and enables us to use $(\mathcal{N};0,1)$ and $(\mathcal{N};0)$ when we wish afterwards to consider only two or one of the constants named, respectively.
\begin{prop}
$\sHom(\mathcal{N};0,1,2)$ is \NP-complete.
\end{prop}
\begin{pf}
Membership in \NP\ is straightforward. Hardness is by reduction from $3$-\textsc{Colouring}. Let $\mathcal{G}$ be an input graph for $3$-\textsc{Colouring}. Let $\mathcal{G}'$ be a structure on signature $\langle R,0,1,2\rangle$,  where $R$ is ternary, built from $\mathcal{G}$ in the following manner. Firstly, add three new elements corresponding to $0$, $1$ and $2$. For each edge $(x,y)$ in $\mathcal{G}$ we introduce three new elements $t_0$, $t_1$ and $t_2$. The constraints $R(0,1,t_2)$, $R(t_2,0,x)$ and $R(t_2,1,y)$ enforce that $(x,y)=(2,2)$ is forbidden. Similarly, $R(1,2,t_0)$, $R(t_0,1,x)$, $R(t_0,2,y)$ and $R(0,2,t_1)$, $R(t_1,0,x)$, $R(t_1,2,y)$ enforce that $(x,y)=(0,0)$ and $(x,y)=(1,1)$, respectively, are forbidden. One may verify that all remaining assignments to $(x,y)$ are attainable. We claim that $\mathcal{G}$ was $3$-colourable iff $\mathcal{G}' \in \sHom(\mathcal{N};0,1,2)$. The forward direction is immediate from the properties of the added constrained elements (we are surjective because of the constants $0$, $1$ and $2$). Likewise, in the backward direction, a (surjective) homomorphism from $\mathcal{G}'$ induces a $3$-colouring of $\mathcal{G}$. 
\end{pf}
\noindent $\Hom(\mathcal{N})$, $\Hom(\mathcal{N};0)$ and $\Hom(\mathcal{N};0,1)$ are all in \Ptime\ (one may colour vertices with just $0$ and $1$, as constrained). It is also easy to see that $\sHom(\mathcal{N})$ and $\sHom(\mathcal{N};0)$  are equivalent (essentially the same problem), due to the symmetry of the template. We already know that $\sHom(\mathcal{N};0,1,2)$ is \NP-complete (as this is true of $\Hom(\mathcal{N};0,1,2)$). The remaining case is $\sHom(\mathcal{N};0,1)$, and this gives us an opportunity to explore the use of \emph{safe gadgets} in reductions. In a typical reduction from $\Hom(\mathcal{A})$ to $\Hom(\mathcal{B})$ one might attempt to simulate the relations of $\mathcal{A}$ through the usage of \emph{gadgets} over $\mathcal{B}$ (this is equivalent to defining the relations of $\mathcal{A}$ over $\mathcal{B}$ using just existential quantification, conjunction and equality). The problem with reductions to $\sHom(\mathcal{B})$ is that this method often fails, as the extra elements in the gadget (corresponding to existential quantification in the definition) typically can make surjective a map that otherwise might not have been. It follows that gadgets that add elements to the problem instance can not in general be used. Safe gadgets enable the addition of extra elements in the gadget because we ensure that these elements can only attain values in the domain that are in any case attained by the elements that were already there. The following proof makes use of such a safe gadget.
\begin{prop}
$\sHom(\mathcal{N};0,1)$ is \NP-complete.
\end{prop}
\begin{pf}
Membership in \NP\ is straightforward. Hardness is by reduction from $\Ret(\mathcal{N}_3)$. Given an instance $I$ of $\Ret(\mathcal{N})$, assume that $u$ is the vertex assigned to $2$. We first remove the preassignment for $2$ in $I$. Then, for each vertex $v \in I$ we add the constraint depicted in Figure~\ref{fig:pcspn}, on new vertices, and identify $x$ with $v$ and $y$ with $u$, and the vertices at the bottom with the vertices in $I$ that are assigned to $0$ and to $1$.
If the resulting instance $I'$ has a surjective solution, then some variable must have been assigned the value $2$. 
First note that we can assume that this variable must be one of the variables from the original variables in $I$, since the right inner variable in Figure~\ref{fig:pcspn} -- $r$ -- can be mapped to $0$ and $1$ only (due to the bottommost constraint), and if the left inner variable -- $l$ -- is mapped to $2$, then $y$ is necessarily mapped to $2$ as well, by case distinction of the value of $r$.

Let $v$ be the variable in $I$ that is mapped to $2$. We claim that in this case the variable $y$ that was introduced for $u$, and therefore also the variable $u$, must also have value $2$. The variable $r$ must be mapped to $1$ because of the rightmost and the lowest grey constraint. Then, because of the topmost grey and the topmost blank constraint, the left variable $l$ must be mapped to $2$. Because of the leftmost grey and another blank constraint, $y$ must be
mapped to $2$. Hence, every surjective solution to $I'$ restricted to the original variables is a correctly preassigned solution to $I$.

Now, consider a solution to the preassigned instance $I$. We know that $y$ is mapped to $2$. If $x$ is also mapped to $2$, we saw above how to satisfy the additional constraints in $I'$. If $x=0$ or $x=1$, we can satisfy the additional constraints by mapping the inner vertices $l$ and $r$ to $0$. Hence, the new instance $I'$ has a surjective solution if and only if $I$ has a solution. Clearly, $I'$ is of polynomial size.
\end{pf}
We introduce another relative of the $k$-no-rainbow-colouring problem that asks for a surjective map that colours each $k$-tuple either with the same colour or with all different colours. This problem and several other surjective homomorphism problems
have been introduced and studied in the context of infinite-domain constraint satisfaction problems in~\cite{Bodirsky} (there it is shown
that several constraint satisfaction problems for trees
can be reduced to surjective homomorphism problems). Formally, let $\mathcal{M}_k$ be over domain $M_k:=\{0,\ldots,k-1\}$ and with $k$-ary relation 
\[ R^{\mathcal{M}_k}:=\{(0,\ldots,0),\ldots,(k-1,\ldots,k-1)\} \cup \{(x_0,\ldots,x_{k-1})\ :\ x_0,\ldots,x_{k-1} \mbox{ distinct} \}.\]
\noindent A structure is connected iff it cannot be specified as the disjoint union of two other structures. A connected component in a structure is an induced substructure that is connected. Proof of the hardness result in the following again makes use of safe gadgets.
\begin{prop}
$\sHom(\mathcal{M}_k)$ is \NP-complete for $k\geq 4$ and tractable for $k\leq 3$.
\end{prop}
\begin{pf}
For $k=1$ or $k=2$ the problem is trivial.
For larger $k$,
we view an instance of $\sHom(\mathcal{M}_k)$ as a $k$-uniform hypergraph.
If the hypergraph has $k$ components, the problem has
a trivial solution, since we can assign the same colour to all
vertices of one component, and different colours to the $k$ different components, and thereby find a homomorphism that uses each of the $k$ colours.
If the hypergraph has fewer than $k$ components, every
surjective homomorphism has to colour the vertices of some constrained $k$-tuple with $k$ colours.

For $k=3$ we can think of the vertices in the hypergraph as elements denoting values in $Z_3$. Owing to the benign properties of the number $3$, i.e. $0+0+0=$ $1+1+1=$ $2+2+2=$ $1+2+3=0 \bmod 3$, a constraining triple $\{v_0,v_1,v_2\}$ in the instance is considered as an equation $v_0+v_1+v_2 \equiv 0 \mod 3$.
Now we select some constrained triple $\{v_0,v_1,v_2\}$, and set the value
of $v_0$ to $0$, $v_1$ to 1, and $v_2$ to 2, and solve the resulting
equation system, e.g., with Gaussian elimination.
If there is a solution, we find a $3$-colouring that uses all colors.
If there is no solution,
we try the same with a different constrained triple.
Suppose there is a surjective homomorphism. As already mentioned, this
homomorphism colours the vertices of some constrained triple $\{v_0, v_1, v_2\}$ with $3$ colours. By symmetry of the colours, we can assume without loss of generality that the homomorphism maps $v_0$ to $0$, $v_1$ to 1, and $v_2$ to 2.
Since the algorithm will eventually choose this constrained triple, it
finds a surjective homomorphism.

Now we prove that the problem is hard for $k=4$ (for larger $k$ we just use the first $k$ entries of the tuples in $R$ with essentially the same proof).
We reduce $3$-\textsc{Colouring} to $\sHom(\mathcal{M}_4)$.
Let $\mathcal{G}$ be an instance of $3$-\textsc{Colouring} which we will assume to be connected. We define an instance $\mathcal{G}'$ of $\sHom(\mathcal{M}_4)$, defined on the vertices of $\mathcal{G}$ and a polynomial number of additional vertices. In the trivial case that $\mathcal{G}$ contains a single vertex, we let $\mathcal{G}'$ be any satisfiable instance to $\sHom(\mathcal{M}_4)$, say $\mathcal{M}_4$.
Otherwise arbitrarily order the edges $e_1, \dots, e_m$ of $\mathcal{G}$.
Let $(xy, uv)$ be a pair or edges, chosen from among the pairs $(e_i, e_{i+1})$ and $(e_m, e_1)$, and insert the following gadget $\mathcal{S}$, where
 $a_0, \dots, a_9$ are new vertices for each pair of edges in $S'$.
The gadget is illustrated in Figure~\ref{fig:colgadget}.
\begin{align*}
%
\{ & \{a_0, a_1, x, x_3\}, \{a_1, a_2, a_3, a_4\}, \{a_4, u, a_6, v\}, \\
& \{x, a_3, y, a_5\}, \{y, a_5, a_7, a_8\}, \{a_5, a_6, a_8, a_9\}\}
\end{align*}
We can see that if $\alpha(x)=\alpha(y)$ in a solution $\alpha$ of $\mathcal{S}$ (a homomorphism from $\mathcal{S}$ to $\mathcal{M}_4$), then all nodes of the gadget have to be assigned the same values, and $\alpha(u)=\alpha(v)$. Moreover we can exhaustively check that if we assign two different values to $u$ and $v$, then we can still consistently assign any two distinct values to $x$ and $y$ and still extend this mapping to a solution of $\mathcal{S}$.
\begin{figure}
\begin{center}
\begin{picture}(0,0)%
\includegraphics{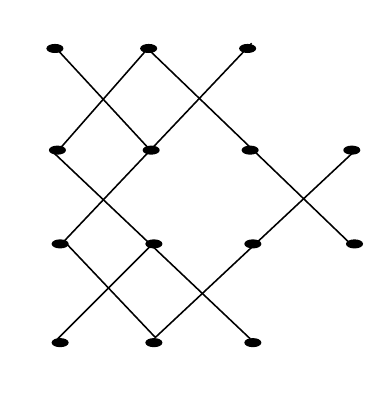}%
\end{picture}%
%
%
\setlength{\unitlength}{3947sp}%
\begingroup\makeatletter\ifx\SetFigFont\undefined%
\gdef\SetFigFont#1#2#3#4#5{%
  \reset@font\fontsize{#1}{#2pt}%
  \fontfamily{#3}\fontseries{#4}\fontshape{#5}%
  \selectfont}%
\fi\endgroup%
\begin{picture}(1805,1885)(224,-1284)
\put(326,-161){\makebox(0,0)[lb]{\smash{{\SetFigFont{10}{12.0}{\rmdefault}{\mddefault}{\updefault}{\color[rgb]{0,0,0}$x$}%
}}}}
\put(326,-624){\makebox(0,0)[lb]{\smash{{\SetFigFont{10}{12.0}{\rmdefault}{\mddefault}{\updefault}{\color[rgb]{0,0,0}$y$}%
}}}}
\put(239,301){\makebox(0,0)[lb]{\smash{{\SetFigFont{10}{12.0}{\rmdefault}{\mddefault}{\updefault}{\color[rgb]{0,0,0}$a_0$}%
}}}}
\put(1014,-174){\makebox(0,0)[lb]{\smash{{\SetFigFont{10}{12.0}{\rmdefault}{\mddefault}{\updefault}{\color[rgb]{0,0,0}$a_3$}%
}}}}
\put(1451,-111){\makebox(0,0)[lb]{\smash{{\SetFigFont{10}{12.0}{\rmdefault}{\mddefault}{\updefault}{\color[rgb]{0,0,0}$a_4$}%
}}}}
\put(1014,-599){\makebox(0,0)[lb]{\smash{{\SetFigFont{10}{12.0}{\rmdefault}{\mddefault}{\updefault}{\color[rgb]{0,0,0}$a_5$}%
}}}}
\put(1489,-674){\makebox(0,0)[lb]{\smash{{\SetFigFont{10}{12.0}{\rmdefault}{\mddefault}{\updefault}{\color[rgb]{0,0,0}$a_6$}%
}}}}
\put(889,-1224){\makebox(0,0)[lb]{\smash{{\SetFigFont{10}{12.0}{\rmdefault}{\mddefault}{\updefault}{\color[rgb]{0,0,0}$a_8$}%
}}}}
\put(1364,-1211){\makebox(0,0)[lb]{\smash{{\SetFigFont{10}{12.0}{\rmdefault}{\mddefault}{\updefault}{\color[rgb]{0,0,0}$a_9$}%
}}}}
\put(876,464){\makebox(0,0)[lb]{\smash{{\SetFigFont{10}{12.0}{\rmdefault}{\mddefault}{\updefault}{\color[rgb]{0,0,0}$a_1$}%
}}}}
\put(1351,451){\makebox(0,0)[lb]{\smash{{\SetFigFont{10}{12.0}{\rmdefault}{\mddefault}{\updefault}{\color[rgb]{0,0,0}$a_2$}%
}}}}
\put(264,-1099){\makebox(0,0)[lb]{\smash{{\SetFigFont{10}{12.0}{\rmdefault}{\mddefault}{\updefault}{\color[rgb]{0,0,0}$a_7$}%
}}}}
\put(1989,-149){\makebox(0,0)[lb]{\smash{{\SetFigFont{10}{12.0}{\rmdefault}{\mddefault}{\updefault}{\color[rgb]{0,0,0}$u$}%
}}}}
\put(2014,-636){\makebox(0,0)[lb]{\smash{{\SetFigFont{10}{12.0}{\rmdefault}{\mddefault}{\updefault}{\color[rgb]{0,0,0}$v$}%
}}}}
\end{picture}%
\end{center}
\caption{The gadget $\mathcal{S}$ for the simulation of 3-colouring in $\sHom(\mathcal{M}_4)$. The $4$-tuples are at the extremities of each ``X''; while they are ordered in this depiction, they may be considered unordered in the reduction.}
\label{fig:colgadget}
\end{figure}
We claim that $\mathcal{G}'$ is a satisfiable instance of $\sHom(\mathcal{M}_4)$ if and only if
$\mathcal{G}$ is 3-colourable. If $\mathcal{G}$ is 3-colourable, we can consistently satisfy all hyperedges in $\mathcal{S}$ according to the above remark and find a surjective homomorphism from $\mathcal{G}'$ to $\mathcal{M}_4$. Now let $\mathcal{G}$ be not 3-colourable, and let $\alpha$ be an arbitrary mapping from $G$ to the three colours. By construction the mapping $\alpha$ corresponds to a partial mapping from $G'$ to $M_k$. We show that this mapping can not be extended to a surjective homomorphism from $\mathcal{G}'$ to $\mathcal{M}_k$. Since $\alpha$ was chosen arbitrary this suffices for the claim. Since $\mathcal{G}$ is not 3-colourable there is an edge $xy$ such that $\alpha(x)=\alpha(y)$. Because the graph $\mathcal{G}$ is connected, and since all of the edges in $\mathcal{G}'$ are strongly connected by the gadget $\mathcal{S}$, all nodes in $\mathcal{G}'$ have to be assigned the same value. Thus no surjective solution exists.
\end{pf}

\section{Final Remarks}

Work in the area of surjective homomorphisms continues to be vigorous at the time of submission. The manuscript \cite{PBBD} contains numerous results (mostly hardness) when the classes of input and template graphs are each restricted (\mbox{i.e.} the template is no longer considered a single graph). In the non-surjective world of MaxCSP, it is known \cite{Creignou,JonssonKK06} that MaxCSP$(\mathcal{B})$ is either in PO or is APX-complete, for $|B|\leq 3$. Zhou \cite{Hang} proves an analog of this in the surjective world -- namely that MaxSurCSP$(\mathcal{B})$ is in PTAS or is APX-complete, for $|B|\leq 3$. Indeed, unlike with MaxCSP$(\mathcal{B})$, there are structures $\mathcal{B}$ such that MaxSurCSP$(\mathcal{B})$ is \NP-hard but has a PTAS.

Finally, we mention that Vikas announces in \cite{Vikas11} his resolution of Questions 1 and 2 (both problems being \NP-complete). No reference is given and the paper containing the proofs is not yet available.

\end{document}